\documentclass[10pt,amsmath,amssymb,graphicx,nofootinbib]{revtex4} 
\usepackage{bm}
\usepackage{graphics,graphicx,calc,epsfig,pstricks,bbm}
\usepackage{amsmath,amssymb,amsfonts}

\newcommand\Tr{{\rm Tr}}

\newcommand{\be}{\begin{equation}}
\newcommand{\ee}{\end{equation}}
\newcommand{\bea}{\begin{eqnarray}}
\newcommand{\eea}{\end{eqnarray}}

\def\extd{\mathrm {d}}

\begin{document}
\title{Beyond Fock space in three dimensional semiclassical gravity}
\author{Michele Arzano}
\email{michele.arzano@roma1.infn.it}
\affiliation{Dipartimento di Fisica and INFN,\\ ``Sapienza" University of Rome,\\ P.le A. Moro 2, 00185 Roma, EU }
\author{Jerzy Kowalski-Glikman}\email{jerzy.kowalski-glikman@ift.uni.wroc.pl}\affiliation{Institute for Theoretical Physics,\\
University of Wroc\l{}aw,\\ Pl.\ Maxa Borna 9, Pl--50-204
Wroc\l{}aw, Poland}
\author{Tomasz Trze\'{s}niewski}\email{tomasz.trzesniewski@ift.uni.wroc.pl}\affiliation{Institute for Theoretical Physics,\\
University of Wroc\l{}aw,\\ Pl.\ Maxa Borna 9, Pl--50-204
Wroc\l{}aw, Poland}

\begin{abstract}
\begin{center}
{\bf Abstract}\\
\end{center}
Quantization of relativistic point particles coupled to three-dimensional Einstein gravity naturally leads to field theories living on the Lorentz group in their momentum representation.  The Lie group structure of momentum space can be traced back to the classical phase space of the particles coupled to topological gravity.  In this work we show how the non-trivial structure of momentum space leads to an unusual description of Fock space.  The latter is reflected in a deformed algebra of creation and annihilation operators which reduces to the ordinary algebra when momentum space ``flattens" to Minkowski space in the limit in which the three-dimensional Newton's constant vanishes.  The construction is covariant under the action of relativistic symmetries acting on the Lorentz group-momentum space.  This shows how it is possible to build a Fock space on a group manifold momentum space in a way consistent with the underlying (deformed) relativistic symmetries.
\end{abstract}

\maketitle
\noindent

\section{Introduction}
The deep power of symmetries in physics is particularly evident in the description of elementary systems in classical and quantum theory.  In quantum field theory for example the very concept of particle is realized in terms of an irreducible representation of the Poincar\'e group, the isometry group of Minkowski space.  This picture is easily understood if we think of the particle as an elementary quantum system for which the action of a Poincar\'e transformation maps a given state into another state of the {\it same} system.  According to the postulates of quantum theory these transformations must be implemented by unitary transformations on a Hilbert space which describes an elementary system if the representation is irreducible.\\
When gravity enters the stage manifolds and metrics much less symmetric are allowed in the game and the space-time will look like Minkowski only locally or in certain asymptotic regions.  Correspondingly in general space-times the notion of particle loses its ``absolute" meaning \cite{Davies:1984rk}.  In three space-time dimensions however solutions to the Einstein's equations will look like Minkowski space {\it everywhere} even though globally they might exhibit non-trivial topologies.  Indeed such lower dimensional incarnation of Einstein's theory does not admit local degrees of freedom and its simplicity makes it a useful toy model for addressing various issues arising in the quest for a theory of quantum gravity.\\
In this work we study some properties of quantum particles coupled to classical gravity in a three dimensional world.  Owing to the topological nature of the theory matter has to be introduced in terms of {\it defects}, in particular particles will be represented as conical defects in an otherwise flat space-time.  Upon quantization we find ourselves in a ``semiclassical" setting where the back-reaction of gravity on the quantum system is forced into the picture in a ``non-perturbative" way via topological nature of the coupling.  This peculiar back-reaction has non-trivial consequences already for the classical description of the particle but its effects become rather dramatic when one considers the associated quantum filed theory.  The main new feature one encounters is that momenta of the particles are parametrized by {\it elements of a Lie group}, in particular of the three-dimensional Lorentz group, rather than by vectors in Minkowski space.  From a field theoretic point of view group valued momenta have appeared in the context of non-commutative field theories associated to deformations of relativistic symmetries almost a decade ago \cite{KowalskiGlikman:2004tz} (for an earlier discussion see \cite{Majid:1994cy}).  It turns out that the quantum field theory emerging from the quantization of topologically gravitating particle is also associated with a certain type of deformed symmetries, the quantum double of the Lorentz group, and that when Fourier transformed to ``space-time" coordinates is indeed a non-commutative field theory (see \cite{Alesci:2011cg} and references therein).  In this work however we focus mainly on the construction of the Hilbert space of the theory in terms of states labelled by momentum eigenvalues.  The reason we focus on this basic aspect is that in analogous four dimensional models a consistent formulation of Fock space has proved to be a surprisingly difficult task \cite{Arzano:2007ef,Daszkiewicz:2007az,Govindarajan:2008qa,Young:2008zm,Arzano:2008bt,Young:2008kf}.  The results we present here show that in three dimensions a formulation of Fock space consistent with the group nature of momenta and their associated deformed symmetries is in fact possible.  The importance of this result is twofold.  On one side it sheds lights on certain structures and mathematical features that will be relevant for addressing the issue of the formulation of the Hilbert space of a wide class of non-commutative field theories whith deformed relativistic symmetries.  More importantly our findings  serve as a invaluable guide in understanding in which direction one should look when searching for features of quantum filed theory which might be modified/altered in the presence of gravitational back-reaction.\\
In the following Section we give detailed arguments of why the momenta of point particles coupled to three dimensional Einstein gravity are parametrized by elements of the Lorentz group.  We do so both from the point of view of Chern-Simons formulation of the theory and in a ``metric" approach.  In Section  III we describe the quantization of our point particle from the classical phase space to the one-particle Hilbert space.  In Section IV we address the issue of constructing multi-particle states with group-valued momenta focusing on the role of the R-matrix and the braid group.  In section V we show how it is possible to generalize the algebra of creation and annihilation operators to the deformed context and we conclude in Section VI with a discussion and outlook for future work.

\section{The momentum space of a topologically gravitating particle}
\subsection{Invitation}
A well known fact about Einstein gravity in 2+1 dimensions is that it does not possess local degrees of freedom \cite{Staruszkiewicz:1963zza}.  Point particles are the simplest example of matter that can be coupled to the theory and due to the topological nature of the latter they must be introduced as {\it defects} surrounded by flat space.  In particular the space-time containing a spinless point particle of mass $m$ is described \cite{Deser:1983tn} by the metric 
\be ds^2 = -d\tau^2 + dr^2 + (1-4Gm)r^2d\varphi^2\,,
\ee
where $G$ is the $2+1$ dimensional Newton's constant with dimension of inverse mass.  The metric above represents a conical space-time, indeed the length of a circular path centered at the origin, the location of the particle, divided by its radius is less than $2\pi$ and the ``deficit angle"  is proportional to the mass of the particle $\alpha=8\pi G m$.  As in the familiar 3+1 dimensional case the mass is a source of curvature however the metric above is locally isometric to Minkowski space and thus the curvature can be only be non-zero at $r=0$, the coordinate singularity and location of the particle.  That this is indeed the case is revealed by calculating the holonomy of a path around the origin which turns out to be a rotation of the deficit angle $\alpha$. \\
As in ordinary $2+1$ Minkowski space we can characterize the physical momentum of the particle, once its mass is given, by specifying two additional parameters which describe the linear momentum and which are in one-to-one correspondence with boosts.  Alternatively we can take the three-momentum of the particle at rest (specified by its mass) and boost it to the appropriate value of the linear momentum.  In this case three-momentum at rest is given by a vector in Minkowski space $\mathbb{R}^{2,1}$.  Since in three dimensions Minkowski space is isomorphic to the Lorentz algebra $\mathfrak{sl}(2,\mathbb{R})$ as a {\it vector space} one can easily see that the physical three-momentum is given by acting with a boost on the vector $mJ_0$, where $J_0$ is the generator of rotation.  In other words the {\it physical} momenta belong to orbits of the adjoint action of the Lorentz group $SL(2,\mathbb{R})$ on its Lie algebra $\mathfrak{sl}(2,\mathbb{R})$. \\
When the particle is described by a conical defect its mass, the
three-momentum at rest, is determined by a {\it rotation} by the
angle $\alpha = 8\pi Gm$ i.e. by $\exp(\alpha J_0) = g_0\in
SL(2,\mathbb{R})$.  The physical momentum can be obtained by
boosting the three momentum at rest, this is now achieved by {\it
conjugating} $g_0$ by a Lorentz boost $L\in SL(2,\mathbb{R})$ \be g
= L^{-1}g_0L\,. \ee Thus the kinematics of a massive particle
coupled to $2+1$-dimensional Einstein gravity described as a moving
conical defect is determined by the set of {\it rotation-like
Lorentz transformations} \cite{Matschull:1997du}.  In analogy with
the usual picture of a point particle in Minkowski space we see that
the {\it extended} momentum space is now given by the group manifold
$SL(2,\mathbb{R})$ (to be contrasted with the vector space
$\mathfrak{sl}(2,\mathbb{R})$ in the ordinary Minkowski case) while
its physical momentum $g = h^{-1}g_0h$ belongs to a given {\it
conjugacy class} of the Lorentz group (to be contrasted with orbits
of the Lorentz group on $\mathfrak{sl}(2,\mathbb{R})$, the
mass-shell, in the Minkowski case).  Below we discuss how the
momentum group manifold picture emerges in a more rigorous setting,
first in the general relativistic picture closer to the heuristic
discussion above and then in the context of the Chern-Simons
formulation of $2+1$-dimensional gravity.

\subsection{Metric approach}
Let us look more in depth at the mechanism for which the effective momentum space of a particle coupled to gravity in $2+1$ dimensions
becomes the Lorentz group manifold using the  language of general relativity \cite{Matschull:1997du}.  We start by imposing a constraint which enforces curvature to be zero everywhere except at the position of the particle, where it acquires a delta like singularity \be\label{2} \frac{1}{2\pi\, G}\, F = L\, mJ_0\, L^{-1}\, \delta^{(2)}(x)\, dx^1\wedge dx^2\,. \ee The solution of the constraint equation (\ref{2}) can be expressed in terms of the dreibein $e = e^aT_a$ and the Lorentz connection $\omega =\omega^aJ_a$, where the Lorentz generators $J_a$ and the translation ones $T_a$ satisfy the known commutation relations \be [J_a,J_b] = \epsilon_{abc}\, J^c\,, \qquad [J_a,T_b] = \epsilon_{abc}\, T^c\,,
\qquad [T_a,T_b] = 0\,. \ee The result is a conical space-time with vanishing torsion and Riemannian curvature equal to zero everywhere
except at the tip of the cone where the particle is placed.  The idea is to interpret this space-time as describing the {\it worldline} of a moving point particle \cite{Matschull:1997du}.  In
order to do so one first has to regularize the curvature singularity.  Let us introduce polar coordinates $(r,\phi)$ on the cone.  The first step will be to cut off the tip of the cone along
the circle of an arbitrary constant $r$ and then cut the resulting surface along the line $\phi = 0$. As a result one obtains a semi-infinite strip, on which again one can introduce the
coordinates $r\in(0,\infty)$, $\phi\in(0,2\pi)$. Since both curvature and torsion on the strip vanish the dreibein and
connection are pure gauge and are given by 
\be\label{12} 
e_\mu = L^{-1}\, \partial_\mu{{\bf q}}\, L\,, \qquad \omega_\mu = L^{-1}\,\partial_\mu L\,, 
\ee 
where $L$ is a scalar field taking values in the group $SL(2,\mathbb{R})$, while ${\bf q}$ is the one taking values in $\mathbb{R}^3$ which is conveniently represented as the vector space of the algebra $\mathfrak{sl}(2,\mathbb{R})$. Both dreibein and connection must be continuous across the cut, i.e. $e_\mu(r,\phi = 0) = e_\mu(r,\phi =2\pi)$, $\omega_\mu(r,\phi = 0) = \omega_\mu(r,\phi = 2\pi)$.  This condition leads to relations for the gauge parameters in (\ref{12})
\be\label{13} 
L_+ = h^{-1}\,L_-\,, \qquad {\bf q}_+ = h^{-1}\left({\bf q}_- - {\bf y}\right)h\,, 
\ee
where the time independent $h$ and ${{\bf y}}$ belong to the Lorentz and translational parts of the Poincar\'e group, respectively and $+/-$ subscripts denote the values at $\phi = 0$ and $\phi = 2\pi$. \\
Since the $r = 0$, $\phi \in (0,2\pi)$ boundary effectively replaces the tip of  the cone it is natural to impose the condition $e_\phi(r
= 0,\phi,t) = 0$ so that the circumference of this boundary vanishes and its time evolution will effectively look like a {\it worldline}.
Denoting the values at $r = 0$ with a bar we see that $\bar{{\bf q}}_+(t) = \bar{{\bf q}}_-(t) \equiv {{\bf x}}(t)$. Then it follows from differentiating (\ref{13})
with respect to time that
the velocity $\dot{{ {\bf x}}} = v^a\, T_a$ is left invariant by the adjoint action of $h$ namely it must commute with such Lorentz group element.  This means that the velocity vector
$\vec{v}$ is parallel to the ``axis" $\vec{p}$ of the Lorentz transformation of the form 
\be
h = p_3\, \mathbf{1} +  p_a\, J^a/\kappa 
\ee
where $\kappa = (4\pi G)^{-1}$ is a constant with dimension of energy.  One naturally identifies $p_a$ with the components of three-momentum of the particle as shown in \cite{Matschull:1997du}. As we saw above the mass of the particle is proportional to the deficit angle of the conical space.  A way to measure the deficit angle is to transport a vector along a closed path around the boundary, as a result it will be rotated by the angle $\alpha = 8\pi Gm$.  Physical momenta will be thus characterized by Lorentz ``holonomies" $h$ which represent a rotation by $\alpha = 8\pi Gm$.  Such requirement imposes the restriction 
\be\label{onshell} 
\frac{1}{2}\mathrm{Tr}(h^2) =
\cos(8\pi Gm)\,\,\,\longrightarrow \,\,\,\, \vec{p}^{\, 2} =
-\frac{\sin^2(4\pi Gm)}{4\pi^2G^2}\,, \ee
on the ``physical" holonomies giving us a ``deformed" mass-shell condition.  From a mathematical point of view such on-shell condition is, as anticipated above, equivalent to imposing that physical holonomies/momenta lie in a given {\it conjugacy class} of the Lorentz group (see e.g. \cite{Schroers:2007ey} for a pedagogical discussion). \\

\subsection{Chern-Simons approach}
The same picture discussed above emerges in the formulation of Einstein gravity in $2+1$ dimensions as a topological field theory.  As shown by Witten \cite{Witten:1988hc} in three space-time dimensios gravity can be formulated as a Chern-Simons theory with the gauge group being (in the case of vanishing cosmological constant) the $2+1$-dimensional Poincar\'e group $ISO(2,1)$.  The latter is a semi-direct product of the groups $SO_{+}(2,1)=SL(2,\mathbb{R})/\mathbb{Z}_{2}$ and $\mathbb{R}^{3}$.
Thus one can decompose a Poincar\'e group element into the Lorentz and translational parts $g = (L,{\bf q})$ and the group multiplication takes the form \be g_1g_2 =
(L_1,{\bf q}_1)(L_2,{\bf q}_2) = (L_1L_2,{\bf q}_1 + \mathrm{Ad}(L_1)\, {\bf q}_2)\,. \ee In order to write down the
Chern-Simons action one needs an invariant inner product on the Lie algebra $\mathfrak{iso}(2,1)$.  It turns out that in $2+1$ dimensions there
are two products available, and it is customary \cite{Witten:1988hc} to choose the one satisfying \be \langle J_a\, T_b\rangle =
\eta_{ab}\,,\qquad \langle J_a\, J_b\rangle = \langle T_a\,
T_b\rangle = 0\,. \ee
The {\it pure} gravity action can be written in terms of the Chern-Simons connection $A = e^aT_a + \omega^aJ_a$.  Using the convenient decomposition $A = A_0\, dt + A_{\cal M}$ with a $\mathfrak{iso}(2,1)$-valued function $A_0$ and a $\mathfrak{iso}(2,1)$-valued one-form $A_{\cal M}$ on the two-dimensional spatial surface ${\cal M}$ we write it in the form
\be\label{a} 
S_{G} = \frac{1}{4\pi G}\int dt\int_{\cal M}\,
\langle\dot A_{\cal M}\wedge A_{\cal M}\rangle + \frac{1}{2\pi
G}\int dt\int_{\cal M}\, \langle A_0\wedge F(A_{\cal M})\rangle\,,
\ee  
where $F(A_{\cal M})$ is the curvature two-form of $A_{\cal M}$.  A point particle can be minimally coupled to the theory \cite{Witten:1988hf}, \cite{deSousaGerbert:1990yp} if
its action is written in terms of a $ISO(2,1)$ element $h = (L,{\bf{\bf q}})$ as \be S_{P} = \int dt\, \langle m\, J_0,\,h^{-1}\dot h\rangle\,. \ee In Minkowski space one can easily see
that this action reproduces the usual action for a relativistic point particle.  Indeed if we write $L\, mJ_0\, L^{-1} ={\bf p}={\bf p}_aJ^a$ and ${\bf {\bf q}} =
{{\bf q}}^aT_a$ and notice that $h^{-1} \dot h = L^{-1}\dot L + L^{-1}\dot{ {\bf q}}\, L$ it is easily shown that 
\be 
S_{P} =\int dt\, {{\bf p}}_a\dot{{{\bf q}}}^a\,, 
\ee as expected.  The full action of the gravity-particle system reads
\be\label{1}
S = \frac{1}{4\pi G}\int dt\int_{\cal M}\, \langle\dot A_{\cal M}\wedge A_{\cal M}\rangle - \int dt\, \langle mJ_0,\, h^{-1}\dot h\rangle\,,
\ee
where the connection $A_{\cal M}$ is constrained by Gauss law (the field equation of $A_0$) according to (\ref{2}) imposed on the curvature $ F(A_{\cal M})$.  The above constraint can be easily solved as follows.  Let us decompose the manifold $\cal M$ into two subregions: the plaquette $\cal D$ being a circle with the center at the position of the particle, on which we introduce coordinates $0 \leq r \leq 1$ and $0 \leq \phi \leq 2\pi$ and the asymptotic region $\Sigma$ with $r \geq 1$.  These two regions have a common boundary $\cal H$, $r = 1$, $0 \leq \phi \leq 2\pi$.  On the asymptotic region the connection is flat and the gauge field takes the form
\be\label{3}
A_{\Sigma} = \gamma\, d\gamma^{-1}\,,
\ee
where $\gamma$ is an element of the Poincar\'e gauge group.  One can find the general solution of (\ref{2}) on the disc as well, it reads
\be\label{4} A_{\cal D} = G\, \bar\gamma\, mJ_0\, \bar\gamma^{-1}\,
d\phi + \bar\gamma d\bar\gamma^{-1}\,,\quad \gamma(0) = h\,. \ee The
fact that (\ref{4}) is a general solution of (\ref{2}) can be easily
checked using the identity $dd\phi = 2\pi\delta(x) dx^1\wedge dx^2$
and that $F(A^g) = gF(A)g^{-1}$, where $A^g$ denotes the gauge
transformed connection and $F(GmJ_0 d\phi) = GmJ_0 dd\phi = 2\pi
GmJ_0\delta(x) dx^1\wedge dx^2$.  In addition we assume that the
gauge field is continuous across the boundary $\cal H$, which
imposes an additional relation between $\gamma$ and $\bar\gamma$
\be\label{5} \left.\gamma\, d\gamma^{-1}\right|_{\cal H} = \left.G\,
\bar\gamma\, mJ_0\, \bar\gamma^{-1}\, d\phi + \bar\gamma\,
d\bar\gamma^{-1}\right|_{\cal H}\,. \ee Decomposing $\gamma = (M,
{{\bf q}})$, $\bar\gamma = (\bar{M}, \bar{{\bf q}})$ a
general solution of (\ref{5}) can be found to read
\cite{Meusburger:2005mg} \be\label{6} M^{-1} = N\,
e^{GmJ_0\phi}\,\bar{M}^{-1}\,,\ \ -\mathrm{Ad}(M^{-1}){{\bf q}}
= {{\bf y}} - \mathrm{Ad}(N\, e^{GmJ_0\phi}M^{-1})\,
\bar{{\bf q}}\,. \ee
where $N$ and ${{\bf y}}$ are generic time-dependent (but space-independent) Lorentz and translation groups elements, respectively.\\
A remarkable thing happens when the explicit forms of the Chern-Simons connection (\ref{3}), (\ref{4}) with the boundary conditions (\ref{6}) are substituted to the action (\ref{1}). Namely the action of the gravity plus particle system collapses to the action describing a deformed particle with curved momentum space being the $SL(2,\mathbb{R})$ group manifold.  Explicitly, consider the {\it group valued momentum} 
\be\label{7} \Pi \equiv L\,
\exp\left(\frac{2}{\kappa}\, mJ_0\right)\, L^{-1} =
\exp\left(\frac{2}{\kappa}\, L\, mJ_0\, L^{-1}\right) = p_3\,
\mathbf{1} + \frac{1}{\kappa}\, p_a\, J^a\,,
\ee 
where $p_3 = \cos({\bf p}/\kappa)$, $p^a =2\kappa{\bf p}^a/{\bf p}\, \sin({\bf p}/\kappa)$.  The ``momentum" $\Pi$ is an element of $SL(2,\mathbb{R})$ so the condition $\label{8} p_3^2 - \frac{1}{4\kappa^2}\, p_ap^a = 1$ must hold, moreover, since the components $p_a$ represent the Lorentz transformed momentum of a particle at rest they satisfy the on-shell relation (\ref{onshell}).  In terms of the group valued momenta $\Pi$ introduced above the the effective action reads
\be\label{10} 
S =- {2}{\kappa}\int dt\, \langle\dot\Pi\Pi^{-1},\,T_a\rangle{\bf q}^a +\int dt\,
\lambda\left(p^2+4\kappa^2\sin^2\frac{m}{\kappa}\right)\,, 
\ee where $\lambda$ is a standard Lagrange multiplier enforcing the mass shell constraint needed if we want to treat all the components $p_a$ as independent variables.\\
Below we illustrate the deep consequences that the Lie group structure of momentum space we just described has in the construction of the quantum one-particle states, the building blocks of the Fock space of the corresponding quantum field theory.

\section{Fields on a conjugacy class: the Hilbert space of a gravitating particle}
The ``one-particle" Hilbert space of a relativistic system is the quantum counterpart of the phase space of a classical relativistic point particle.  As discussed in the Introduction as a {\it relativistic} and {\it elementary} quantum system such space should carry an irreducible representation of the Poincar\'e group (see e.g. \cite{Newton:1949cq}).  In ordinary 3d Minkowski space-time the space of complex functions on the (positive) mass-shell equipped with the invariant inner product
\be
(f,g)=\int d^3 p\, \delta (p^2-m^2) \theta(p_0) \overline{f}(p) g(p)
\ee
is a Hilbert space on which the Poincar\'e group acts irreducibly and indeed is taken as the space of states of a single massive quantum relativistic particle \cite{Weinberg:1995mt}.\\
We look here for a quantum counterpart of the phase space with ``curved" momentum space $SL(2,\mathbb{R})$.  In general given a space, not necessarily a {\it vector} space, on which the action of a group is defined, one can construct a representation of such group on the set of complex functions on the given space \cite{VileK}.  Indeed in a non-gravitational setting Minkowski space naturally carries an action of the Poincar\'e group and so will the complex functions on it.  However the representation of the Poincar\'e group on such space of functions {\it is not} irreducible.  One has to look at the restriction to the set of functions with support on orbits of the Lorentz group and in particular, to describe a massive particle, to the positive mass-shell.\\
As we showed in the previous Section for a relativistic particle coupled to three-dimensional gravity orbits of the Lorentz group on Minkowski space are replaced by conjugacy classes of the three-dimensional Lorentz group, i.e. ``orbits" of the group acting on itself by conjugation.  Thus one is naturally led to consider complex functions on conjugacy classes of $SL(2,\mathbb{R})$ as Hilbert spaces of ``conical" particles with a given mass $m$.  Such intuition is indeed correct as first explored in \cite{Carlip:1989nz} and further shown in \cite{Bais:1998yn}.  In particular in \cite{Carlip:1989nz} Carlip, following the strategy first adopted in \cite{'tHooft:1988yr}, quantized the gravitating particle starting from the classical phase space and using the Chern-Simons formulation of 3d gravity; the resulting Hilbert space is given by functions on a given conjugacy class.\\
In \cite{Bais:2002ye} it was shown that the space of functions on a conjugacy class carries an irreducible representations of a non-trivial Hopf algebra, the {\it quantum double} of the Lorentz group, also known as the ``Lorentz double", \cite{Koornwinder:1996uq}.  The latter can be seen as a deformation of the group algebra of the Poincar\'e group (for more details see \cite{Bais:2002ye}).  Thus, roughly speaking, the effect of gravity on quantum particles in 3d is to {\it deform} the group of relativistic symmetries into a non-trivial Hopf algebra or {\it quantum group}.  A detailed discussion of the role of the Hopf algebra structure of the Lorentz double is beyond the scope of the present work (see \cite{Fuchs:1997jv} for a basic introduction).  Here we would like to remark that the non-trivial {\it co-algebra} structure of these deformed models governs the action of the symmetry generators on tensor products of representations and, for example, is reflected on the non-abelian addition of momenta discussed in the next Section.\\
Our starting point here will be the Hilbert space of a particle with mass $m$ realized in terms of functions on the ``positive energy" subspace of the {\it conjugacy class} given by elements of $SL(2, \mathbb{R})$ conjugate to a rotation by  an angle $8\pi Gm$.  A canonical choice of positive energy subspace is given in terms of the condition $p_0 > 0$ (see \cite{Matschull:1997du}) on the 'cartesian' matrix parametrization of group elements (\ref{7}) given above.  Functions on the Hilbert space are square integrable with respect to the invariant measure
\be
d\mu (g)_{G} \equiv d\mu(g) \delta \left(1/2\,\mathrm{Tr} (g^2) - \cos (8\pi Gm)\right)\,.
\ee
where $d\mu(g)$ is the (left and right-invariant) Haar measure on the group and the Dirac delta has support on the given conjugacy class.  The associated inner product will be given by
 \be\label{innerp}
(\psi,\phi)_{G}=\int \extd \mu (g)_{G}\, \theta(p_0)\, \overline{\psi(g)}\,  \phi(g)\, ,
\ee
where $\theta(p_0)$ restricts the support of the integrand to the ``positive energy" hyperboloid.\\
As customary we label one-particle states in terms of elements of conjugacy classes of $SL(2,\mathbb{R})$ which we denote simply by $|g\rangle$.  Since these kets carry an irreducible representation of the Lorentz double we must have an action of Lorentz transformations and translation generators on them.  The Lorentz group has natural {\it left} and {\it right} actions on itself
\be
g\vartriangleright h \equiv g^{-1} h g\,,\,\,\,\,\,\,\,\,\,\,\,\,h\vartriangleleft g \equiv g h g^{-1}\,,
\ee
and from our discussion above we know that functions on the group will carry a left and right representations of the group so that our kets will transform as
\be
\Lambda_L (g) |h\rangle \equiv  |g^{-1}\vartriangleright h\rangle \equiv |g h g^{-1}\rangle\,,\,\,\,\,\,\,\,\,\,\,\,\, \Lambda_R (g) |h\rangle \equiv  |h \vartriangleleft g^{-1}\rangle \equiv |g^{-1} h g\rangle\,,
\ee
Notice that the two actions are related by group inversion, i.e. $g\vartriangleright_R\, (.) \equiv g^{-1}\vartriangleright_L\, (.)$ and thus it suffices to restrict to one of the two.
Once a choice of group action has been made it is easily checked that it corresponds to the Lorentz sector of the representation of the Lorentz double as constructed in \cite{Bais:1998yn, Koornwinder:1996uq}.  From these representations one can also read off the action of the translation generators which are formally the same as ordinary translations i.e.
\be
T({\bf x})\vartriangleright |h\rangle \equiv e^{\frac{1}{2\kappa} \Tr (h {\bf x})} \vartriangleright |h\rangle 
\ee
where ${\bf x}= {\bf x}_a J^a\in \mathfrak{sl}(2,\mathbb{R})$.  The analogy with the unitary action of ordinary translations in three-dimensional Minkowski space is manifest in the parametrization of $h\in SL(2,\mathbb{R})$ given by ``cartesian" coordinates ${\bf p}^a(h)=\frac{\kappa}{2} \Tr (h J^a)$ for which
\be
T({\bf x})\vartriangleright |h\rangle \equiv e^{i {\bf p}^a(h){\bf x}_a} |h\rangle\, .
\ee
In this representation we can write down ``translation generators"\footnote{Notice that there is no unique choice for such generators since any other parametrization of group elements could in principle be used to define a new {\it basis} of translation generators.  This ambiguity is well known in models with group manifold momentum space \cite{Arzano:2010jw}.} whose eigenvalues give us the momentum carried by a one-particle state
\be
P^a \vartriangleright |h\rangle \equiv {\bf p}^a(h) |h\rangle\, ,
\ee
which will be useful for what follows.  The scalar product (\ref{innerp}) above becomes the following relation for bra and kets in terms of (generalized) functions on the conjugacy class
\be
\langle h | g \rangle = \delta_{h} (g)\,.
\ee
Note that the delta function which appears above is invariant under conjugation $\delta_h (g) = \delta (h^{-1} g) = \delta (k^{-1} h^{-1} g k) = \delta_{k^{-1}hk} (k^{-1}g k)$.  In closing this section we notice that in the limit $G\rightarrow 0$ the conjugacy class ``flattens out" into an orbit of the Lorentz group in Minkowski space and we re-obtain the one particle Hilbert of ordinary quantum field theory in terms of functions on the positive energy mass-shell square integrable w.r.t. the ordinary Lebesgue measure.

\section{Multi-particle states: deformed symmetrization}
As seen above quantum states representing a single massive scalar particle are labelled by elements of conjugacy classes of $SL(2,\mathbb{R})$.  We now move to the description of the multiparticle sector of the space of states of the theory. In ordinary (bosonic) Fock space constructions \cite{Geroch:1985ci}  indistinguishability of elementary quantum particles requires that states which describe more than ``one particle" are described by symmetrized tensor products of one-particle states.  This simple procedure becomes quite non-trivial in the present context.\\  To see this let us consider the total momentum carried by a simple tensor product of one particle states $|h_1\rangle\otimes|h_2\rangle$.  The action of a translation generator on such tensor product is determined by the ``co-algebra" structure of the Lorentz double \cite{Bais:1998yn, Alesci:2011cg} and it is intuitively quite simple
\be P^a \vartriangleright
(|h_1\rangle\otimes|h_2\rangle) = {\bf p}^a (h_1h_2)(|h_1\rangle\otimes|h_2\rangle)\,. 
\ee 
The essential feature of the action above is that due to the non-abelian nature of the momentum group manifold the two tensor product states $|h_1\rangle\otimes|h_2\rangle$ and $|h_2\rangle\otimes|h_1\rangle$ will have {\it different total momenta}: ${\bf p}^a(h_1h_2)\neq {\bf p}^a(h_2 h_1)$.  In particular the ``ordinary" symmetrized state
$\frac1{\sqrt{2}}\,\left(|h_1\rangle\otimes|h_2\rangle + |h_2\rangle\otimes|h_1\rangle\right)$ {\it is not} an eigenstate of the translation generators.  If we interpret pictorially the flip of tensor products in terms of an ``exchange" of particles we see that by a local switch of particles we have altered a global property of the system, the total momentum, which is defined by a holonomy around both defects.  If we assume instead that when a particle moves around the other it gets conjugated by the holonomy of the first then the local exchange will not affect the global momentum of the system.  Indeed a two-particle state given by 
\be 
|h_1h_2\rangle
\equiv \frac1{\sqrt{2}}\,\left(|h_1\rangle\otimes|h_2\rangle +
|h_1h_2h_1^{-1}\rangle\otimes|h_1\rangle\right) 
\ee 
has well defined momenta ${\bf p}^a(h_1h_2)$, as it can be easily checked.  Notice that we can make the same ansatz but keeping the particle labelled
by $h_1$ fixed and having the other going around it.  We can thus write down what we call ``left" and ``right" symmetrization 
\bea
|h_1h_2\rangle_L & \equiv & \frac1{\sqrt{2}}\,\left(|h_1\rangle\otimes|h_2\rangle + |h_1h_2 h_1^{-1}\rangle\otimes|h_1\rangle\right)\,, \nonumber\\
|h_1h_2\rangle_R & \equiv & \frac1{\sqrt{2}}\,\left(|h_1\rangle\otimes|h_2\rangle + |h_2\rangle\otimes|h_2^{-1}h_1h_2\rangle\right)\,. 
\eea
As we will see below the ``left" and ``right" labels are justified by the relation of the different symmetrizations with the adoption of left or right action of the Lorentz group on itself.  Let us remark here that, not surprisingly, the fact that the momentum of a particle gets conjugated by the holonomy of the other particle when ``exchanging" the two has strict analogy with the phenomenon of {\it flux metamorphosis} in lower dimensional gauge theories in presence of excitations carrying topological charges \cite{Bais:1980vd, Lo:1993hp}. \\
Before moving on in the description of our construction let us discuss the covariance properties of the proposed symmetrization.  As we have seen in the previous Section the Lorentz group acts on the kets by conjugation.  On tensor products of representations the action of the Lorentz group should be read off the {\it co-product} of the Lorentz double and it turns out \cite{Koornwinder:1996uq} that such action is {\it group-like} and thus it is the usual action that ordinary Lorentz transformations would have on tensor product states namely
\be
\Delta\Lambda_{L,R}(g) |h_1\rangle\otimes|h_2\rangle \equiv \Lambda_{L,R}(g) |h_1\rangle\otimes\Lambda_{L,R}(g) |h_2\rangle\,.
\ee
It is straightforward to check that both {\it left} and {\it right} symmetrizations above are {\it covariant} under such transformations namely
\bea
(\Lambda_L(g)\otimes\Lambda_L(g)) |h_1,h_2\rangle_{L} & \equiv & |g h_1g^{-1},g h_2 g^{-1}\rangle_{L}\,, \nonumber\\
(\Lambda_R(g)\otimes\Lambda_R(g)) |h_1,h_2\rangle_{R} & \equiv & |g^{-1} h_1 g, g^{-1} h_2 g\rangle_{R}\,.
\eea
This is reassuring since the deformed symmetrization we propose is not in conflict with the underlying symmetries of the system.  At a closer look this should not come as a surprise since it can be shown that the Lorentz double (as any other quantum double construction) possesses a quantum $R$-matrix from which one can construct a generalization of the usual ``flip" operator ($\sigma(a\otimes b) = b\otimes a$) which acts as an intertwiner of tensor product representations.  Indeed for the Lorentz double the $R$-matrix acts on a tensor product of kets as the transformation
\be
R \equiv \sum_{g\in G} \delta(g^{-1}.)\mathbbm{1}\otimes\Lambda_{L}(g)
\ee
and the $R_{21}$-matrix, which is its reverse-order variant, acts as
\be
R_{21} \equiv \sum_{g\in G} \Lambda_{R}(g)\otimes\delta(g^{-1}.)\mathbbm{1}\,.
\ee
If we now assume a left action of the Lorentz group on kets one can easily show that 
\be |h_1h_2\rangle_L \equiv
\frac1{\sqrt{2}}\,\left(\mathbbm{1}\otimes\mathbbm{1} + \sigma\circ
R\right) |h_1\rangle\otimes|h_2\rangle\,. 
\ee 
On the other hand setting the action of Lorentz generators on kets as the right action
one has 
\be 
|h_1h_2\rangle_R \equiv
\frac1{\sqrt{2}}\,\left(\mathbbm{1}\otimes\mathbbm{1} + \sigma\circ
R_{21}\right) |h_1\rangle\otimes|h_2\rangle\,. 
\ee
It is straightforward to show that the above prescription for symmetrization extends to a generic $n$-particle state.  Indeed in this case one can construct all possible permutations of a given $n$-fold tensor product of one-particle states using, instead of the ordinary exchange operator $\sigma_i (|h_1\rangle\otimes\ldots|h_i\rangle\otimes
|h_{i+1}\rangle\ldots\otimes|h_n\rangle) = |h_1\rangle\otimes\ldots|h_{i+1}\rangle\otimes
|h_i\rangle\ldots\otimes|h_n\rangle$, the new operator $\tau_L(i) \equiv (\sigma\circ R)_i$, whose action reads
\be
\tau_L(i) (|h_1\rangle\otimes\ldots|h_i\rangle\otimes
|h_{i+1}\rangle\ldots\otimes|h_n\rangle) = |h_1\rangle\otimes\ldots|h_ih_{i+1}h^{-1}_i\rangle\otimes
|h_i\rangle\ldots\otimes|h_n\rangle
\ee
and analogously for the {\it right} symmetrization.  Notice that the exchange operator does not square to the identity and thus its action can be seen as a representation of the {\it braid group} \cite{Kauffman:1991ds} rather than the symmetric group, obeying the corresponding relations
\bea
\tau_{L,R}(i)\tau_{L,R}(j) & = & \tau_{L,R}(j)\tau_{L,R}(i)\,,\ \ |i-j|\geq 2 \,, \nonumber\\
\tau_{L,R}(i)\tau_{L,R}(i+1)\tau_{L,R}(i) & = & \tau_{L,R}(i+1)\tau_{L,R}(i)\tau_{L,R}(i+1)\,.
\eea
Explicitly a given $n$-particle state can be built iteratively:
\bea
\vert h_{1}\ldots h_{n+1}\rangle_{L} = \mathcal{S}_{L}(n+1)\left(\vert h_{1}\rangle\otimes\vert h_{2}\ldots h_{n+1}\rangle_{L}\right)\,, \nonumber\\
\mathcal{S}_{L}(n+1) \equiv
\frac{1}{\sqrt{n+1}}\Big(\mathbbm{1}^{\otimes(n+1)} +
\sum_{i=1}^{n}\left(\mathbbm{1}^{\otimes(i-1)}\otimes(\sigma\circ
R)\otimes\mathbbm{1}^{\otimes(n-i)}\right)\circ\ldots\circ\left((\sigma\circ
R)\otimes\mathbbm{1}^{\otimes(n-1)}\right)\Big) \eea 
and analogously for the {\it right} symmetrization.  Let us notice at this point that using higher powers of the exchange operator $\tau_L =\sigma\circ R$ one can construct several candidates for multiparticle states.  For example the two-particle state 
\be
\frac1{\sqrt{2}}\,\left(|h_1 h_2 h_1^{-1}\rangle\otimes|h_1\rangle +
|h_1h_2h_1h_2^{-1}h_1^{-1}\rangle\otimes|h_1h_2h_1^{-1}\rangle\right)\,,
\ee 
obtained by ``braiding" the state $|h_1h_2h_1^{-1}\rangle\otimes|h_1\rangle$ is an eigenstate of the momentum operator with eigenvalue ${\bf p}^a(h_1h_2)$ like the state 
\be
\frac1{\sqrt{2}}\,\left(|h_1\rangle\otimes|h_2\rangle +
|h_1h_2h_1^{-1}\rangle\right)\,. 
\ee 
It can be noted however that
\be 
|h_1h_2h_1^{-1}\rangle\otimes|h_1\rangle +
|h_1h_2h_1h_2^{-1}h_1^{-1}\rangle\otimes|h_1h_2h_1^{-1}\rangle =
\Delta\Lambda(h_1) \left(|h_2\rangle\otimes|h_1\rangle +
|h_2h_1h_2^{-1}\rangle\otimes|h_2\rangle\right)\,. 
\ee 
Similarly braiding the state $|h_1h_2h_1h_2^{-1}h_1^{-1}\rangle\otimes|h_1h_2h_1^{-1}\rangle$ one obtains a state with total momentum ${\bf p}^a(h_1h_2)$ which is nothing but $\Delta\Lambda(h_1h_2) (|h_1h_2\rangle_L)$.  In other words candidate two-particle states obtained using multiple powers of the exchange operator can be obtained by Lorentz transforming ``basic" two-particle states constructed using just one power of $\sigma\circ R$ as long as both {\it left} and {\it right} symmetrization are considered within the physical Fock space.

\section{Braiding commutators: the proposal}
In this Section we explore the possibility of defining analogues of creation and annihilation operators which starting from a vacuum state $|0\rangle$ defined by
\be
a_{L,R}(h)|0\rangle = 0\,,\ \ a_{L,R}^{\dagger}(h)|0\rangle = |h\rangle
\ee
will span the full Fock space of the theory.  As noticed above Lorentz transformations by one of the labels of the basis kets act non-trivially in multiparticle states.  In particular the following property holds
\bea\label{RLsymm}
(\Delta \Lambda_L(h_1^{-1})) \triangleright |h_1h_2\rangle_L & = & |h_2h_1\rangle_R\,, \nonumber\\
(\Delta \Lambda_L(h_2)) \triangleright |h_1h_2\rangle_R & = & |h_2h_1\rangle_L\,,
\eea
which shows that there is a non-trivial relation between left- and right-symmetrized states and their counterparts with inverted labels: they are connected by a Lorentz transformation involving one of the ``momentum" group elements.  One can go further and prove the following useful relation
\be
(\Lambda_L(h_2)\circ\Lambda_L(h_1^{-1})\otimes\mathbbm{1}) |h_1h_2\rangle_L = |h_2 h_1\rangle_L\,.
\ee
Defining $\mathcal{L}_L(h_1, h_2) \equiv (\Lambda_L(h_2)\circ\Lambda_L(h_1^{-1})\otimes\mathbbm{1})$ we can then write the following {\it braided} commutators
\bea
a^{\dagger}_L(h_1)a^{\dagger}_L(h_2) - \mathcal{L}^{-1}_L(h_1, h_2) a_L^{\dagger}(h_2)a_L^{\dagger}(h_1) & = & 0\,, \nonumber\\
a_L(h_1)a_L(h_2) - a_L(h_2)a_L(h_1) \mathcal{L}_L(h_1,h_2) & = & 0
\,, \eea 
where by definition $a_L^{\dagger}(h_1)a_L^{\dagger}(h_2) |0\rangle \equiv
|h_1h_2\rangle_L$.  While the commutators written above in terms of Lorentz transformations give a more ``physical" picture of the deformed creation and annihilation operators, from the algebraic point of view it would be more satisfactory to re-write the above commutators in terms of the $R$-matrix.
This is indeed possible but the outcome is not more illuminating that the relations above and reads
\bea
a_L^{\dagger}(h_2)a_L^{\dagger}(h_2^{-1}h_1h_2) - (\sigma\circ R_{21}^{-1}) a_L^{\dagger}(h_1)a_L^{\dagger}(h_2) & = & 0\,, \nonumber\\
a_L(h_2^{-1}h_1h_2)a_L(h_2) - a_L(h_2)a_L(h_1) (R_{21}^{-1}\circ\sigma) & = & 0\,,
\eea
where it is intended that $R_{21}^{-1} \equiv \sum_{h \in G} \Lambda_L(h^{-1})\otimes\delta(h^{-1}.)\mathbbm{1}$ and $R^{-1} \equiv \sum_{h \in G} \delta(h^{-1}.)\mathbbm{1}\otimes\Lambda_R(h^{-1})$.  
One could now proceed in an analogous way starting from the right action of the Lorentz group on one-particle states and define {\it right} operators which create and annihilate right-symmetrized states.  However since the latter are connected by Lorentz transformations to the left symmetrized states as shown in (\ref{RLsymm}) the explicit expressions for the left-braided commutators given above will be sufficient for our illustrative purposes. \\
To find the ``cross"-commutators between $a_L(h)$ and $a_L^{\dagger}(h)$ we use the inner product $\langle h|g\rangle = \delta(h^{-1}g)$ and the invariance of the delta function under conjugation $\delta(h^{-1}g) = \delta(k^{-1}h^{-1}gk)$.  Then we consider the action of an annihilation operator on a two-particle state
\be
a_L(h_1)(a_{L}^{\dagger}(h_2)a_{L}^{\dagger}(g) |0\rangle) = \langle h_1|h_2\rangle |g\rangle + \langle h_1|h_2gh_2^{-1}\rangle |h_2\rangle = \delta(h_1^{-1}h_2) |g\rangle + \delta(h_1^{-1}h_2 g h_2^{-1}) |h_2\rangle\,.
\ee
Now using the invariance of the delta function under conjugation we see that
\be
\delta(h_1^{-1}h_2 g h_2^{-1}) |h_2\rangle \equiv \delta(h_2^{-1}(h_1^{-1}h_2 g h_2^{-1}) h_2) |h_2\rangle = \delta(h_2^{-1}h _1^{-1} h_2 g) |h_2\rangle \equiv a_{L}^{\dagger}(h_2)a_L(h_2^{-1}h_1h_2) |h\rangle\,,
\ee
from which we can infer the commutator
\be
a_L(h_1)a_L^{\dagger}(h_2) - a_{L}^{\dagger}(h_2)a_L(h_2^{-1}h_1h_2) = \delta(h_1^{-1}h_2)\,.
\ee
Of course one can proceed in a similar way to obtain an analogous expression for the {\it right} operator.  Our conjecture is that the ``braided" commutators obtained above using only the properties of two-particle states are the correct generalization of the algebra of creation and annihilation operators of ordinary quantum field theory for our ``braided" field theory model.  It remains however to be proved that these results extend to a generic $n$-particle states and such task is left for future investigation.

\section{Discussion}
The main lesson we learn in attempting to construct the Hilbert space of a relativistic particle gravitating in 2+1 dimensions is that its defect-like nature profoundly affects the multiparticle structure of the theory.  We found that the non-abelian composition of group-valued momenta labelling one-particle states imposes a {\it braided symmetrization} in the construction of multiparticle states.  This reflects a non-trivial algebraic structure underlying the symmetries of the particle's phase space governed by the {\it quantum double} of the Lorentz group whose mathematical aspects were studied in the past \cite{Koornwinder:1996uq} but which here are applied for the first time in a field theoretic context.  Under the appropriate generalization of Fourier transform our field defined on a group manifold can be mapped into a {\it non-commutative} field defined on the corresponding Lie algebra.  The analysis presented thus provides the first consistent {\it covariant} deformed/braided Fock space for a non-commutative quantum field theory of this type.  Field theories exhibiting similar mathematical structures in 3+1 dimensions have been widely studied in the past but the construction of a covariant Fock space compatible with the deformed symmetries has been a challenging problem which remains open.  We believe that our results will be of guidance for solving this long standing issue.  Finally this work aims to show in a clear and straightforward way how certain field theories beyond ordinary local QFT naturally emerge in a very simple combination of Einstein gravity and quantum theory in lower dimensions.  This kind of ``toy-semiclassical gravity" shows that QFT undergoes deep structural changes in the presence of topological gravity \cite{Arzano:2012bj} and may indicate potential avenues for analogous modifications in four dimensions where the connection of similar non-commutative field theories to (quantum) gravity is far from being an established fact.

\begin{acknowledgments}
MA work is supported by EU Marie Curie Actions through a Career Integration Grant and in part by a grant from the John Templeton Foundation.  JKG is supported in part by the grant 2011/01/B/ST2/03354.  JKG and TT are supported by funds provided by the National Science Center under the agreement DEC-2011/02/A/ST2/00294. TT acknowledges the support by the Foundation for Polish Science International PhD Projects Programme co-financed by the EU European Regional Development Fund and the support by the European Human Capital Program.
\end{acknowledgments}


\begin{thebibliography}{99}

\bibitem{Davies:1984rk}
  P.~C.~W.~Davies,
  ``Particles Do Not Exist,''
  In *Christensen, S.m. ( Ed.): Quantum Theory Of Gravity*, 66-77

\bibitem{KowalskiGlikman:2004tz} 
  J.~Kowalski-Glikman and S.~Nowak,
  hep-th/0411154.
  
\bibitem{Majid:1994cy} 
  S.~Majid and H.~Ruegg,
  Phys.\ Lett.\ B {\bf 334}, 348 (1994)
  [hep-th/9405107].  

\bibitem{Alesci:2011cg}
  E.~Alesci and M.~Arzano,
  Phys.\ Lett.\ B {\bf 707}, 272 (2012)
  [arXiv:1108.1507 [gr-qc]].

\bibitem{Arzano:2007ef}
  M.~Arzano and A.~Marciano,
  Phys.\ Rev.\ D {\bf 76}, 125005 (2007)
  [arXiv:0707.1329 [hep-th]].

\bibitem{Daszkiewicz:2007az}
  M.~Daszkiewicz, J.~Lukierski and M.~Woronowicz,
  Mod.\ Phys.\ Lett.\ A {\bf 23} (2008) 653
  [hep-th/0703200].

\bibitem{Govindarajan:2008qa}
  T.~R.~Govindarajan, K.~S.~Gupta, E.~Harikumar, S.~Meljanac and D.~Meljanac,
  Phys.\ Rev.\ D {\bf 77}, 105010 (2008)
  [arXiv:0802.1576 [hep-th]].

\bibitem{Young:2008zm}
  C.~A.~S.~Young and R.~Zegers,
  Nucl.\ Phys.\ B {\bf 809}, 439 (2009)
  [arXiv:0807.2745 [hep-th]].

\bibitem{Arzano:2008bt}
  M.~Arzano and D.~Benedetti,
  Int.\ J.\ Mod.\ Phys.\ A {\bf 24}, 4623 (2009)
  [arXiv:0809.0889 [hep-th]].

\bibitem{Young:2008kf}
  C.~A.~S.~Young and R.~Zegers,
  Commun.\ Math.\ Phys.\  {\bf 298}, 585 (2010)
  [arXiv:0812.3257 [math.QA]].


\bibitem{Staruszkiewicz:1963zza}
  A.~Staruszkiewicz,
  Acta Phys.\ Polon.\  {\bf 24}, 735 (1963).

\bibitem{Deser:1983tn}
  S.~Deser, R.~Jackiw and G.~'t Hooft,
  Annals Phys.\  {\bf 152}, 220 (1984).

\bibitem{Matschull:1997du}
  H.~J.~Matschull and M.~Welling,
  Class.\ Quant.\ Grav.\  {\bf 15}, 2981 (1998)
  [arXiv:gr-qc/9708054].

\bibitem{Schroers:2007ey}
  B.~J.~Schroers,
  PoS QG {\bf -PH}, 035 (2007)
  [arXiv:0710.5844 [gr-qc]].

\bibitem{Witten:1988hc}
  E.~Witten,
  Nucl.\ Phys.\  B {\bf 311} (1988) 46.

\bibitem{Witten:1988hf}
  E.~Witten,
  Commun.\ Math.\ Phys.\  {\bf 121} (1989) 351.

\bibitem{deSousaGerbert:1990yp}
  P.~de Sousa Gerbert,
  Nucl.\ Phys.\ B {\bf 346} (1990) 440.

\bibitem{Meusburger:2005mg}
  C.~Meusburger and B.~J.~Schroers,
  Nucl.\ Phys.\ B {\bf 738} (2006) 425
  [hep-th/0505143].


\bibitem{Newton:1949cq}
  T.~D.~Newton and E.~P.~Wigner,
  Rev.\ Mod.\ Phys.\  {\bf 21}, 400 (1949).

\bibitem{Weinberg:1995mt}
  S.~Weinberg,
  ``The Quantum theory of fields. Vol. 1: Foundations,''
  Cambridge, UK: Univ. Pr. (1995) 609 p

\bibitem{VileK}
N.J.~Vilenkin and A.U.~Klimyk, ``Representation of Lie Groups and
Special Functions,'' 
3~volumes Kluwer, 1991, 1993.



\bibitem{Carlip:1989nz}
  S.~Carlip,
  Nucl.\ Phys.\ B {\bf 324}, 106 (1989).

\bibitem{Bais:1998yn}
  F.~A.~Bais and N.~M.~Muller,
  Nucl.\ Phys.\ B {\bf 530} (1998) 349
  [hep-th/9804130].


\bibitem{'tHooft:1988yr}
  G.~'t Hooft,
  Commun.\ Math.\ Phys.\  {\bf 117}, 685 (1988).

\bibitem{Bais:2002ye}
  F.~A.~Bais, N.~M.~Muller and B.~J.~Schroers,
  Nucl.\ Phys.\ B {\bf 640}, 3 (2002)
  [hep-th/0205021].

\bibitem{Koornwinder:1996uq}
  T.~H.~Koornwinder and N.~M.~Muller,
  J. Lie Theory {\bf 7} 33-52 (1997)
  [arXiv: q-alg/9605044].

\bibitem{Fuchs:1997jv}
  J.~Fuchs and C.~Schweigert,
  ``Symmetries, Lie algebras and representations''
  Cambridge, UK: Univ. Pr. (1997) 438 p
  
  
\bibitem{Arzano:2010jw} 
  M.~Arzano,
  Phys.\ Rev.\ D {\bf 83}, 025025 (2011)
  [arXiv:1009.1097 [hep-th]].

  \bibitem{Geroch:1985ci}
  R.~P.~Geroch,
  ``Mathematical physics,''
  The University of Chicago Press, 1985

\bibitem{Bais:1980vd}
  F.~A.~Bais,
  Nucl.\ Phys.\ B {\bf 170}, 32 (1980).

\bibitem{Lo:1993hp}
  H.~-K.~Lo and J.~Preskill,
  Phys.\ Rev.\ D {\bf 48}, 4821 (1993)
  [hep-th/9306006].

\bibitem{Kauffman:1991ds}
  L.~H.~Kauffman,
  Singapore, World Scientific (1991) 538 p. (Series on knots and everything,~1)


\bibitem{Sasai:2007me}
  Y.~Sasai and N.~Sasakura,
  Prog.\ Theor.\ Phys.\  {\bf 118} (2007) 785
  [arXiv:0704.0822 [hep-th]].


\bibitem{AmelinoCamelia:2007zzb}
  G.~Amelino-Camelia, A.~Marciano and M.~Arzano,
  In Di Domenico, A. (ed.): Handbook of neutral kaon interferometry at a Phi-factory, 155-186

\bibitem{Arzano:2012bj}
  M.~Arzano,
  arXiv:1212.1097 [hep-th].




\end{thebibliography}
\end{document}